# Magnetic properties in a partially oxidized nanocomposite of Cu-CuCl

Qi Li[1], Shi-Wei Zhang[1,3], Yan Zhang[2] and Chinping Chen[2,4]

[1] State Key Laboratory for Structural Chemistry of Unstable and Stable Species, College of Chemistry and Molecular Engineering, Peking University, Beijing 100871, P. R. China

[2] Department of Physics, Peking University, Beijing 100871, P. R. China

E-mail : cpchen@pku.edu.cn

**Abstract.** Magnetism of a very thin antiferromagnetic (AFM) surface CuO has been investigated with the partially oxidized nanocomposites of Cu-CuCl, $\sim 200$ nm. The samples are characterized by X-ray diffraction, X-ray photoelectron spectroscopy, X-ray-excited Auger electron spectroscopy, transmission electron microscope and magnetic measurements. The characterizations indicate that the composites have a core-shell structure. Before the oxidation, it is $(Cu)_{core}/(CuCl)_{shell}$, and after the oxidation, $(Cu)_{core}/(Cu_2O+CuCl+minuteCuO)_{shell}$. The magnetic measurements have revealed that a ferromagnetic (FM) like open hysteresis exists at the temperature below the freezing point, $T_F$. In the high field region, a paramagnetic (PM) response appears without showing a sign of saturation. Also, the field dependent magnetization ($M$-$H$) measurement is PM-like at $T > T_F$. These interesting magnetic properties are evident to arise from the AFM CuO on the outer surface. They are attributed to the uncompensated surface spins of $Cu^{2+}$ and the effect of surface random potential. More interestingly, the magnetic susceptibility is greatly enhanced in the presence of $Cl^-$ anions at $T < T_F$, according to the field-cooled/zero-field-cooled (FC/ZFC) measurements. This further supports the point that the disorder or frustration effect of the impurity would reduce the AFM ordering of CuO and increase the level of uncompensated spins.
Submitted to Nanotechnology

## 1. Introduction

The magnetic properties of mesoscopic or nanoscaled material have become increasingly important due to the application potential in many areas, such as ultra-high density magnetic recording [1], spintronics, etc [2]. With the progresses in technology, various shapes of the nanoscaled magnetic materials, such as the nanorods, nanowires, *etc*., have been fabricated or synthesized. To understand the nanomagnetics of these materials is, therefore, a critical issue not only for the fundamental interest but also for the practical purposes [3-5]. With nanoscaled particles (NPs), the investigation on their magnetism has long been one of the focused points [6, 7]. In particular, many studies have been reported on the magnetic properties of AFM NPs, including NiO [8-10], CoO [11-13], ferritin [14, 15], $\alpha$-$Fe_2O_3$ [16, 17] and ferrihydrites.[18, 19]. There are several properties observed with the AFM NPs which are not yet fully understood [20, 21]. Among them, the substantially large moment and the FM-like magnetism associated with the AFM NPs are the interesting points for further studies [9].

---

[3] zhangsw@pku.edu.cn,
[4] Phone : +86-10-62751751, Fax : +86-10-62751615

The surface effect is very important to the magnetic behaviour of NPs due to the existence of surface random potential or the surface anisotropy. In a couple of previous experiments [9, 22], a core-shell model emphasizing the surface effect has been applied to explain the magnetic properties associated with the NPs. The surface spin glass state would result in a marked effect on the magnetism not only for the ferro- or ferri- NPs, but also for the AFM NPs. A very recent experiment has demonstrated that a very thin layer, ~ 1 nm, of AFM CoO surrounding a FM Co core does not exhibit an exchange biased behaviour, whereas such effect has been observed with a thicker CoO layer, ~ 3 nm [13]. This indicates that for a CoO layer to show AFM property, the thickness has to exceed 1 nm. Similar result has been reported in another experiment with $Co_{core}/CoO_{shell}$ NPs having the core-shell structure as well [23]. Although, there is still in need of a satisfactory explanation on the underlying mechanism, it has already demonstrated the importance of surface effect. In addition, the size dependent magnetic moment of NiO NPs has been studied to show that the surface spin coordination has a profound effect on the multiplicity of the AFM sublattice configuration in the whole particle. The irreversibility in the spin state between different configurations would result in a large coercivity and a shifted loop [9].

Recently, some unique magnetic properties of CuO NPs have been reported [24, 25], which are attributed to the variation in unit cell and coordination style of surface copper with respect to oxygen. In this paper, we report a detailed investigation on the magnetic properties of partially oxidized, diamagnetic nanocomposites of Cu-CuCl. After the oxidation, the NPs are characterized to show a core-shell structure, $(Cu)_{core}/(CuCl+Cu_2O+minute\ CuO)_{shell}$. Since the Cu, CuCl, and $Cu_2O$ are all diamagnetic, the observed magnetism is reasonably inferred to arise from the minute amount of AFM CuO residing over the surface. Therefore, the sample offers a good model system to study the magnetism of surface AFM CuO.

## 2. Synthesis and measurements

A typical sample was prepared by the following procedure. $CuCl_2 \cdot 6H_2O$ in water solution was reduced to metal Cu NPs at 80 $^o$C by hydrazine hydrate. Afterwards, the precipitates of Cu NPs were washed several times by water to remove the unreacted hydrazine hydrate in the solution. A small amount of $CuCl_2 \cdot 6H_2O$ was then added again to form a CuCl shell enclosing the Cu NPs. Then, the Cu-CuCl composites were filtered out, washed three times by ethanol, and dried in vacuum. The final partially oxidized product was obtained through controlled air oxidation of the Cu-CuCl composite powder at room temperature after various days of oxidation, from 6 to 42 days. During the process of synthesis, the CuCl shell is critical in order to obtain a sample with interesting magnetic properties, although the molar ratio of CuCl in the shell over the Cu in the core accounts for only 1 %.

The samples thus obtained were characterized by several techniques. The crystal phase was analyzed by X-Ray diffraction (XRD) using a Rigaku diffractometer and CuKα radiation of wavelength, λ = 0.15418 nm. The elements chemical state on the surface of the Cu-CuCl composites before and after the oxidation was analyzed by X-ray photoelectron spectroscopy (XPS) and X-ray-excited Auger spectrum (XAES). The XPS spectra were recorded using an Axis Ultra spectrometer (Kratos, UK) with monochromatic AlKα radiation (1486.71 eV) at a power of 225 W (15 mA, 15 kV). To compensate for the surface charge effects, binding energies $E_b$ were calibrated using C 1s hydrocarbon peak at 284.8 eV. The XAES was performed on the same spectrometer, Axis Ultra spectrometer with the same AlKα radiation source. The structure of the nanocomposites was further investigated by a Hitachi H-9000 transmission electron microscope (TEM). Magnetic properties of these nanocomposite powders were measured using Quantum Design SQUID magnetometer.

## 3. Results

The rechults of aracterization for the sample phases, compositions, and structure, by XRD, XPS, XAES, and TEM are presented. The magnetic properties and the corresponding analyses are given as well.

### 3.1. XRD analysis on Cu, Cu-CuCl and partially oxidized Cu-CuCl NPs

The XRD patterns for the Cu metal and the Cu-CuCl nanocomposites before and after 30 days of oxidation are presented in figure 1. The grain size (D) was calculated from the width of the XRD peaks using the Debye-Scherrer relation after correcting for the instrumental broadening. It indicates that the composites are made up of cubic Cu (D ≈ 26.7 nm), cubic $Cu_2O$ (D ≈ 12.6 nm) and cubic CuCl (D ≈ 24.8 nm). A noteworthy point is that, only the peak for $Cu_2O$ shows up after the oxidation, while, there is not any detectable CuO within the detection sensitivity of XRD. However, the characterization by XPS and XAES evidences the presence of $Cu^{2+}$.

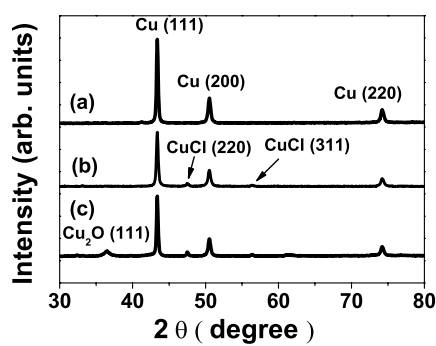

Figure 1, XRD patterns for (a) the Cu metal, (b) Cu-CuCl before oxidation, and (c) partially oxidized Cu-CuCl. In the curve for the oxidized sample, no CuO phase is detected, except for the peak of $Cu_2O$.

### 3.2. Evidences of $Cu^{2+}$ with the partially oxidized sample by XPS and XAES

The XPS curves for the Cu-CuCl nanocomposites before and after 30 days of oxidation are shown in figure 2. The curve for the sample before oxidation displays two symmetric peaks corresponding to the Cu $2p_{3/2}$ (932.6 eV) and Cu $2p_{1/2}$ (952.5 eV). On the other hand, in the curve for the partially oxidized sample, additional shoulders show up at 934.6 eV and 954.7 eV, $i.e.$, higher $E_b$ value than in the main peaks. These are owing to the presence of CuO [26]. Furthermore, the accompanied shake-up satellite peaks approximately 9 eV higher than the shoulders are a signature for the $Cu^{2+}$. Therefore, the XPS analysis confirms that CuO appears after the oxidation, and exists on the surface of the NPs. In addition, XAES for the oxidized sample is recorded and shown in the inset of figure 2. The peak maximum of Cu $L_3VV$ is at a kinetic energy of 916.3 eV ($E_b$ ~ 570.2 eV), agreeing with the literature $Cu_2O$ value. However, the peak shape is closer to that of CuO, confirming the presence of $Cu^{2+}$ on the surface [27]. From these analyses with XRD, XPS, and XAES, the composites should be a complex, which has a core of Cu, a shell of $Cu_2O$ and CuCl. In addition, a little CuO possibly exists on the outer surface.

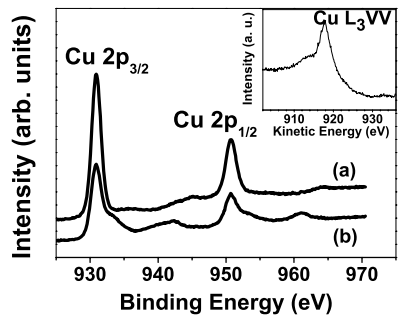

Figure 2, XPS for the Cu-CuCl (a) before and (b) after 30 days of oxidation. Inset is the Cu $L_3$VV of XAES for the oxidized sample. The peak position agrees with the literature value of $Cu_2O$, while the peak shape is closer to that of CuO.

### 3.3. Structure analysis by TEM

The structure of these composites was further characterized with TEM. Figure 3a shows the TEM images of the Cu NPs before the oxidation. The shape is close to spherical with a uniformly dark shade and a sharp edge, indicating the uniformity in the composition. The particle size, ~ 200 nm, is much larger than the grain size calculated from XRD, indicating the crystal structure is polycrystalline within the NPs. TEM images of the partially oxidized Cu-CuCl nanocomposites are shown in figure 3b. It reveals the core-shell structure of the nanocomposites. The core appears with a much darker contrast than that of the shell, due to the difference in electron penetration efficiency. The shell appears to be fuzzy in a much lighter shade, almost the same as that of the background in the photo, but is still discernable. The shell contains diamagnetic compositions of $Cu_2O$ and CuCl with a little of AFM CuO. Its thickness is estimated to be 10 - 20 nm.

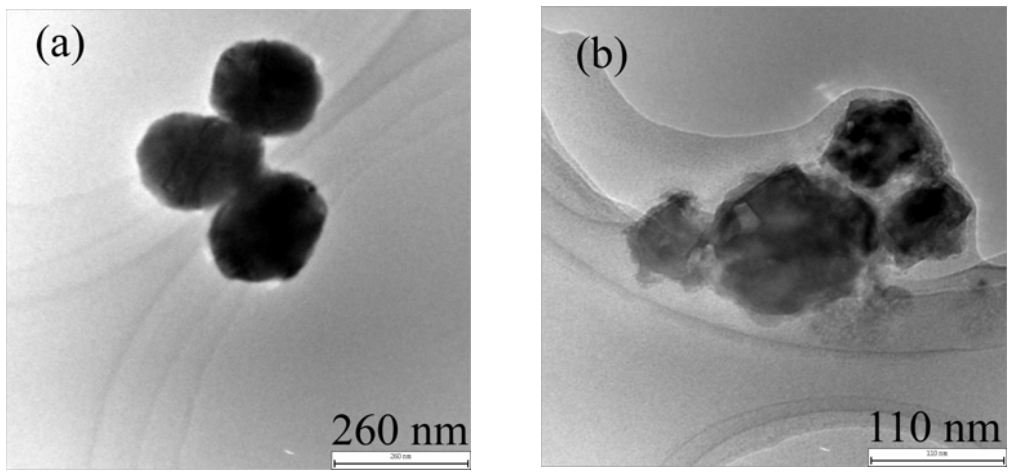

Figure 3, TEM images of (a) metal Cu nanoparticles and (b) partially oxidized Cu-CuCl nanocomposites. A shell structure in a lighter shade surrounding a dark core is visible..

### 3.4.Temperature dependent magnetic susceptibility of partially oxidized Cu-CuCl nanocomposites

Temperature-dependent field-cooled (FC) and zero-field-cooled (ZFC) magnetic susceptibility, $\chi = M/H$, of the composite after 30 days of oxidation is shown in figure 4. The FC and ZFC curves exhibit a typical freezing behaviour at a temperature below the freezing temperature, $T_F$ ~ 6.4 K, which is determined by the maximum in the ZFC curve. At $T < T_F$, $\chi_{FC}$ increases rapidly with decreasing temperature and becomes ~ 17 times the value of $\chi_{ZFC}$. As mentioned above, the major composition of the sample is diamagnetic Cu, CuCl and $Cu_2O$, which do not exhibit large magnetic moments. Hence, the origin of the observed magnetism at $T < T_F$ with $\chi_{FC}$ is from the small amount of CuO on the surface. At about 8 K, which is above $T_F$, the FC and ZFC curves collapse. The susceptibility follows well-behaved Curie-Weiss law, $\chi = C/(T-\theta)$, where C is a constant. The inset (a) in figure 4 gives $\theta = -24.6$ K for the sample after 30 days of oxidation. The Curie-Weiss parameter, $\theta$, indicates the AFM nature of the spin interaction in the samples. In addition, $\chi_{FC}$ for the samples after 6, 12, 18, 24, 30, 36, and 42 days of oxidation has been measured. The samples with longer oxidation period exhibit larger magnetic susceptibility at $T < T_F$. The inset (b) in figure 4 shows this behaviour by the $\chi_{FC}$ curves for the samples after 6, 24, 36, 42 days of oxidation. The larger $\chi_{FC}$ with the sample of longer oxidation period should be attributed to the higher $Cu^{2+}$ contents. The FC and ZFC magnetic susceptibility of the pellet and the powder samples after 30 days of oxidation were also measured as a function of temperature, as shown in figure 5. The results show that $\chi$ of the pellet sample is much smaller than that of the powder, especially at low temperature. Since the powder sample has a much larger open surface than that of the pellet, it further confirms that the magnetic moments of these composites are arising primarily from the surface.

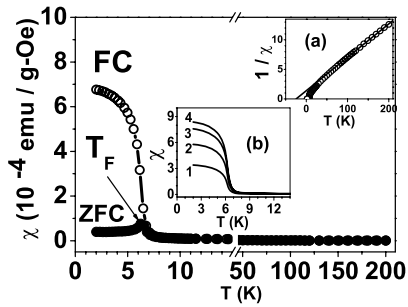

Figure 4, Temperature variation of $\chi$ under ZFC and FC at 100 Oe field for the sample after 30 days of oxidation. $T_F \sim 6.4$ K. Inset (a) shows that $\chi$ follows a well-behaved Curie-Weiss law at T > 70 K. Inset (b) shows $\chi$ under FC for the samples after the various oxidation periods, (1) 6 days, (2) 24 days, (3) 36 days, (4) 42 days.

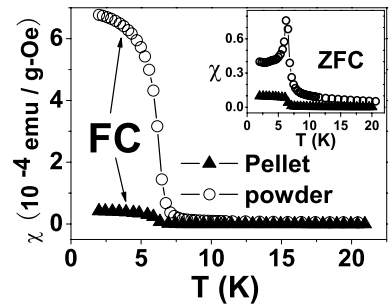

Figure 5, Temperature variation of $\chi$ for the pellet and powder under ZFC and FC at 90 Oe field. The powder has higher $\chi$ by almost an order of magnitude at $T < T_F$.

### 3.5. Effect of $Cl^-$ anions on magnetic property

Figure 6 shows the FC susceptibilities for the samples with and without CuCl. Both samples are after 30 days of oxidation. The sample with CuCl exhibits much higher value of $\chi_{FC}$ at $T < T_F$. At $T = 2$ K, it is ~16 times the value for the sample without CuCl. This suggests that the presence of $Cl^-$ anions tends to enhance the magnetism. The inset shows the ZFC data for the two samples. Likewise, the sample with $Cl^-$ has a much pronounced susceptibility than the sample without $Cl^-$. At $T = T_F$, the difference grows to be an order of magnitude.

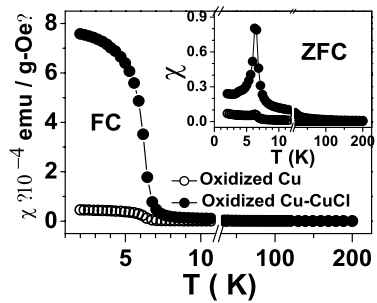

Figure 6, Temperature variations of $\chi_{FC}$ with and without CuCl at 100 Oe. The oxidized Cu NPs were prepared with the same oxidation period (30 days) as the one of oxidized Cu-CuCl. The inset is for $\chi_{ZC}$.

*3.6.Field dependent magnetization of partially oxidized Cu-CuClnanocomposites*

The field dependent magnetization ($M$–$H$) curves for $(Cu)_{core}/(Cu_2O+CuCl+minuteCuO)_{shell}$ after 30 days of oxidation were measured at 2 K, 6 K and 10 K, and presented in figure 7. At $T$ = 2 K and 6 K, *i.e.* below $T_F$, open hysteresis loops are observed,. It evidences a FM-like behaviour. The remanent magnetization and coercive field at $T < T_F$ are 0.124 emu/g, 5225 Oe at 2 K and 0.065 emu/g, 508 Oe at 6 K. The open loops close at about 25 kOe and 2.1 kOe measured at 2 K and 6 K, respectively. The much enhanced coercivity and remanence at 2 K in comparison with those at 6 K, clearly shown in the inset, is strongly correlated to $\Delta M = M_{FC} - M_{ZFC}$ at the same temperatures, see figure 4 for the $M_{FC}$ and $M_{ZFC}$. $\Delta M \sim 6.4$ emu/g at 2 K, and 2.6 emu/g at 6 K. This indicates that the open hysteresis at $T < T_F$ is attributed to the irreversibility of the surface moment, since $\Delta M$ is arising from the surface magnetism. Above $T_F$, the $M$–$H$ curve reveals a linear PM behaviour. This is consistent with the result of FC and ZFC measurements, showing $\Delta M = 0$ at $T > T_F$. FC hysteresis loop was measured, too, at 2 K in order to investigate whether an exchange-biased effect exists. A field of 30 kOe was applied during the cooling process from 300 K down to 2 K. No such effect has been observed.

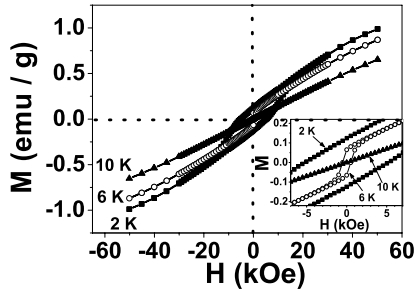

Figure 7, Hysteresis loops of the powder sample at 2K, 6K and 10K with the cooling history of ZFC. The inset is for the loops in the low-field region.

## 4. **Discussion**

The magnetic measurements along with the characterizations on the structure and composition have provided strong evidences that the observed magnetism is arising from the surface. This indicates that the FM-like open hysteresis loop at $T < T_F$ is attributed to the thin layer of AFM CuO over the surface surrounding the nano-sized diamagnetic core. More interestingly, the existence of the Cl⁻ anions would greatly enhance the magnetism below $T_F$.

In some of the previously reported experiments, the magnetization hysteresis loops have been observed with the nickel ferrite $(NiFe_2O_4)$ [22], NiO [9], and CuO [24] nanoparticles. The FM-like irreversibility has been interpreted as due to the transition between the multiple surface spin configurations resulting from the surface disorder state, which influences the magnetization reversal of the whole particle. In our experiment, the magnetism is a pure surface effect without the interference from the background magnetism of the core. So, indeed, the surface effect plays an important role to cause the FM-like behaviour. The surface moment is expected to arise from the uncompensated surface spins of the AFM CuO.

The PM response in the high field region of the $M$-$H$ curve measured at $T < T_F$, and the PM behaviour with the $M$-$H$ curve measured at $T = 10$ K, see figure 7, can be understood in terms of the same surface uncompensated spins as discussed above. Similar PM response in the high field region has been observed with the above-mentioned NiO [9] and CuO [24] NPs, and in particular, with the Co(core)/CoO(shell) samples by Tracy et al. [13]. In this analogy, the authors have ascribed it to the existence of defect moment in the CoO shell. However, according to our experiment, the existence of the Cl⁻ impurity actually enhances the magnetism at $T < T_F$.

## 5. Conclusions

Complex Cu-CuCl composites with a core-shell structure have been synthesized. They consist of a Cu core, a shell of $Cu_2O$ and CuCl, along with a little CuO on the outer surface. The magnetism from a thin surface layer of AFM CuO contributes a substantial value of the total magnetic moment. Below the freezing temperature, $T_F$, it shows a FM-like open hysteresis along with a PM response in the high field region. In the meanwhile, a PM-typed *M-H* curve has been observed at $T > T_F$. The observed magnetic moment on the particle's surface can be explained by the uncompensated surface spin. In addition, the presence of $Cl^-$ would greatly enhance the magnetism, possibly by adding the surface disorder. The samples are unique in the magnetic properties of the AFM NPs and are helpful in understanding the nature of the magnetism in such particles.

Acknowledgement

This work was supported by NSFC of China (No. 20273001).